# Explaining Counts from EPRB Experiments: Are They Consistent with Quantum Theory?

James H. Bigelow
jlandjhb@gmail.com
December 14, 2011

## ABSTRACT

I have fit data from EPRB experiments of Weihs et al to a model that consists of an EPRB thought experiment whose output is filtered. The filter implements the assumptions made by most investigators: that EPRB experiments satisfy a fair sampling assumption; and that detections and coincidences occur randomly with Poisson distributions (this provides a basis for calculating standard errors of coincidence counts, correlations, and so forth). The model does not fit the data—predicted and observed counts of detections and coincidences differ far too much, by a chi-square criterion. Logically, one must give up fair sampling and/or Poisson errors and/or the assumption that the data derive ultimately from an EPRB thought experiment.

In the literature, giving up fair sampling seems to be coupled to giving up EPRB, but to me it seems just as sensible to keep EPRB and to give up fair sampling and Poisson errors. Some rather ordinary mechanisms can violate fair sampling, and there is the possibility in any experiment for uncontrolled and unmonitored factors to contribute to unwanted variation. By sufficiently relaxing the fair sampling and Poisson assumptions—it doesn't take much—my model can be made to fit the data.

## INTRODUCTION

In 1997 and 1998, Weihs et al performed a series of experiments (Weihs et al, 1998; Weihs, 2007) to test whether nature can violate a Bell inequality[1] (Bell, 1964). This paper examines one set of 41 of those experiments performed on May 1, 1998 and labeled *scanblue110* through *scanblue151*.[2] The two observers, conventionally labeled Alice and Bob, were also given the colors blue (for Alice) and red (for Bob). Thus "scanblue" means that Alice's measurement settings—two in each experiment, separated by 45°—scanned through a range of angles, while Bob employed the same two measurement settings in all 41 experiments.

---

[1] The term "Bell inequality" refers to any of a large number of inequalities that classical physics says must be obeyed by the correlations found in nature, but that quantum theory claims can be violated. Bell (1964) found the first one. Clauser et al (1969) discovered another. There are quite general methods for producing Bell inequalities by the bushel basket (e.g., Avis et al, 2005; Bigelow, 2008; Peres, 1999).
[2] The labels *scanblue110 – scanblue151* would suggest there were 42 experiments. But *scanblue138* was a copy of *scanblue137*, so I omitted it.

Experiments to test for Bell inequality violations using correlated pairs of either photons or spin-1/2 massive particles are called EPRB experiments because the ability of entanglement to cause "spooky action at a distance" was first pointed out in the famous 1935 paper by Einstein, Podolsky, and Rosen; and David Bohm (the "B" in "EPRB") was the first to suggest the use of spin-1/2 particles to test this spooky prediction of quantum theory (Bohm, 1951).

The possibility of a Bell violation is only one prediction of quantum theory, and it is logically possible that the experiments disagree with some other quantum theory predictions. This paper is primarily an attempt to determine whether the *scanblue* experimental data are consistent with quantum theory in general. I am not the first to ask this question. Adenier and Khrennikov (2006, 2007) examined the *scanblue* data, and concluded that something weirder even than quantum theory was needed to explain it. De Raedt et al (2012) analyzed data from 23 of the experiments of Weihs et al—not the *scanblue* experiments, but very similar—and concluded that the data were inconsistent with quantum theory. I question assumptions made in both papers, hence this re-examination.

Here is my approach. I assume, tentatively, that the *scanblue* data were generated by an EPRB thought experiment, and then filtered. The filtering process can both attenuate and distort the signal from the thought experiment. It should be local-realistic (not something weirder than quantum theory), physically justified (there should be footprints of it in the data), and as simple as possible. I will start with the simplest of filters, and discover that it doesn't do the job. I'll try a sequence of ever more elaborate filtering mechanisms, each fitting the data better than the last. If the sequence leads to a model that fits the data well enough (by a criterion to be introduced later), then I will conclude that the *scanblue* data are consistent with quantum theory.

This paper does not investigate whether the data could have been generated by an entirely classical process (local realist, if you prefer). I include this disclaimer because the latter question is currently a subject of lively debate. Readers could well expect it to figure heavily in a paper on EPRB experiments. The fact that I do not attempt to fit a local realist model to the data does not mean that I believe it can't be done. Nor does it mean I believe it can be done. This could be a worthy subject for a later paper, but it has nothing to do with this one.

## THE *SCANBLUE* EXPERIMENTS

In this section I briefly describe the experiments performed by Weihs et al and the data collected from them. A more complete discussion can be found in Weihs (2007).

As shown in Fig. 1 (figures are at the end of the paper), two observers (Alice and Bob) each have clock and a box. Each box is connected by an optical fiber to a central device. Each observer applies one of two voltages to her box, switching between them at random times, independently of the other observer.

Every so often a box emits a pulse on one of two channels. Alice and Bob accumulate logs of their pulses, including: (1) the time of the pulse; (2) which of the two voltages was being applied at that time (called the *setting* and coded as 0 or 1); and which of the two channels emitted the pulse (called the *result* and also coded as 0 or 1). In each *scanblue* experiment, Alice's log contains about 200,000 and Bob's about 140,000 detections, collected during a five second interval.

It is assumed that these pulses are caused by events occurring at the central device—i.e., that they are detections of photons transmitted from the central device via the connecting fibers—so that if a pair of pulses, one logged by Alice and one by Bob, occur at about the same time, then they have a common cause. Pairs with a common cause are called *coincidences*. Identifying them in the data involves a good bit of analysis (Bigelow, 2009). Only a few percent of the logged pulses are paired in this way.

Let subscripts $i, j, k, l$ index Alice's setting, Alice's result, Bob's setting, and Bob's result, respectively, all taking values 0 or 1. Let the superscript $m$ index the 41 *scanblue* experiments. $m$ takes the values *scanblue110*, *scanblue111*, …, *scanblue151* (missing *scanblue138*, as explained in an earlier footnote). From each experiment $m$, therefore, I obtain 24 counts, four *singles* for Alice $a_{ij}^m$, four singles for Bob $b_{kl}^m$, and sixteen coincidences $c_{ijkl}^m$. This comes to a total of 984 counts for the 41 *scanblue* experiments.

It will be convenient to define *unpaired* singles as:

$$ua_{ij}^m = a_{ij}^m - \sum_{kl} c_{ijkl}^m \qquad ub_{kl}^m = b_{kl}^m - \sum_{ij} c_{ijkl}^m$$

Every pulse logged by Alice and Bob is either an unpaired single or is included in one and only one of the coincidences.

## A NOTE ON NOTATION

I give my 'Alice' variables the name 'a', and I adorn them with sub- and superscripts, accent marks, etc., to identify which variable in the 'Alice' class I am talking about. Similarly, I give variables in the 'Bob' class the name 'b', and those in the 'Coincidence' class the name 'c'. Unaccented variables are counts from the actual data. A 'hat' over a variable means it is a predicted value, 'bar' means it is an average. You will encounter 'hats' and 'bars' later.

At this point I ran out of places to put distinguishing marks, so I started adding another letter as a prefix. Do not interpret two-letter names as products of two one-letter variables. Prefix 'u' means unpaired, 'p' means detection probability (identification probability for coincidences), 'q' means quantum probability (i.e., theoretical probability calculated from the EPRB recipe). You have already seen the prefix 'u'. The others will occur shortly.

## THE EPRB THOUGHT EXPERIMENT

Two correlated photons are generated and distributed to two observers (this is the function of the central device and connecting fibers in Fig. 1). Their state is represented by a 4×4 density matrix (i.e., a self-adjoint positive semi-definite matrix with trace 1), which I denote as $\rho$. Unlike most papers that discuss EPRB experiments, I will not take $\rho$ to be the singlet state. Instead, for each of my models I will choose the density matrix along with other free parameters to best fit the model to the *scanblue* data. Among other things, then, this paper offers a method to perform quantum tomography.

Measurements by Alice or Bob are represented by 2×2 self-adjoint operators. For each experiment $m$, Alice has four 2×2 operators $A_{ij}^m$, one for each setting/result combination $(i,j)$. Let $I$ be the 2×2 identity operator. Using the trace rule, the probability $qa_{ij}^m$ that Alice obtains result $j$ given that she performs measurement $i$ in experiment $m$ is:

(1a) $$qa_{ij}^m = Tr\left((A_{ij}^m \otimes I)\rho\right)$$

Similarly, Bob will have four 2×2 operators $B_{kl}^m$, corresponding his setting/result combinations $(k,l)$.[3] The probability $qb_{kl}^m$ that Bob obtains result $l$ given that he performs measurement $k$ in experiment $m$ is:

(1b) $$qb_{kl}^m = Tr\left((I \otimes B_{kl}^m)\rho\right)$$

Finally, the probability $qc_{ijkl}^m$ that Alice obtains result $j$ and Bob obtains result $l$ given that Alice performs her measurement $i$ and Bob his measurement $k$ in experiment $m$ is:

(1c) $$qc_{ijkl}^m = Tr\left((A_{ij}^m \otimes B_{kl}^m)\rho\right)$$

Appendix A describes the operators for all 41 *scanblue* experiments.

## ASSESSING THE GOODNESS OF FIT

My goodness-of-fit measure is the standard chi-square statistic (e.g., see Brownlee, 1960, Chap 5).

Each of the models I develop will provide predictions of the 984 counts. Denote the predictions of such a model as $\hat{a}_{ij}^m, \hat{b}_{kl}^m, \hat{c}_{ijkl}^m$. Denote the predictions of unpaired singles as:

$$\widehat{ua}_{ij}^m = \hat{a}_{ij}^m - \sum_{kl} \hat{c}_{ijkl}^m \qquad \widehat{ub}_{kl}^m = \hat{b}_{kl}^m - \sum_{ij} \hat{c}_{ijkl}^m$$

To measure how well these predictions match the data, I need estimates of the variances of the counts. It is usual to assume that each count has a Poisson distribution, for which the variance equals the mean. If the expected value of the count is not too small (larger than ten, as a rough rule of thumb), then a Normal distribution is a good approximation, with the same mean and variance as the Poisson. On the assumption that the model's predictions are correct, $(ua_{ij}^m - \widehat{ua}_{ij}^m)^2 / \widehat{ua}_{ij}^m$, $(ub_{kl}^m - \widehat{ub}_{klj}^m)^2 / \widehat{ub}_{klj}^m$, and $(c_{ijkl}^m - \hat{c}_{ijkl}^m)^2 / \hat{c}_{ijkl}^m$ are all squares of standard Normal random variables, i.e., chi-square with one degree of freedom, and they are all mutually independent (it is to achieve independence that I take unpaired singles rather than total singles).

---

[3] Since Bob uses the same two settings in every experiment, his operators are independent of the experiment $m$. I retain the superscript $m$ anyway, simply for symmetry.

Suppose a model $M$ has a number $F$ of free parameters (possibly a different $F$ for each model, of course) which I will adjust to achieve the best possible fit of the model to the data. Then the expression:

$$(2) \quad X(M) = \sum_m \left[ \sum_{ij} \frac{(ua_{ij}^m - \widehat{ua}_{ij}^m)^2}{\widehat{ua}_{ij}^m} + \sum_{kl} \frac{(ub_{kl}^m - \widehat{ub}_{kl}^m)^2}{\widehat{ub}_{kl}^m} + \sum_{ijkl} \frac{(c_{ijkl}^m - \hat{c}_{ijkl}^m)^2}{\hat{c}_{ijkl}^m} \right]$$

will have a chi-square distribution with $DF = 984 - F$ degrees of freedom. Since $984 - F$ will be a fairly large number, the distribution of $X(M)$ will be approximately Normal with mean $DF$ and variance $2 \times DF$. If $X(M)$ is many standard deviations above its mean (e.g., if $Z(M) = (X(M) - DF(M))/\sqrt{2 \times DF(M)}$ is five or larger), then model $M$ leaves too much of the variation in the *scanblue* data unexplained, and I will consider the model to be inconsistent with the data.[4]

A less demanding criterion might well be justified. Surely there are variations in conditions from one *scanblue* experiment to the next, conditions that are imperfectly controlled and imperfectly known. This ignorance can be represented as randomness; in classical physics, ignorance is the excuse for treating anything as random. I will defer further discussion of relaxing the goodness-of-fit criterion until I have established a need to do so.

## THE MODELS

In his "Bertlmann's socks" paper, Bell (1981) points out: "It is true that practical [EPRB] experiments fall far short of the ideal, because of counter inefficiencies, or geometrical imperfections, and so on. It is only with added assumptions, or conventional allowance for inefficiencies and extrapolation from the real to the ideal, that one can say the [Bell] inequality is violated."

In each of the models described below, I define a purely classical, local-realistic mechanism—a filter—that attenuates and distorts the ideal quantum probabilities (1a-c). One can think of this filter, operating in reverse, as extrapolating a la Bell from the real observed counts to the ideal quantum probabilities. The behavior of the filter depends on some parameters, mostly detection probabilities, that I can adjust to fit the model's predictions to the observed counts. As mentioned earlier, I also adjust the density matrix elements. The values of the parameters that I choose are those that minimize the model's X statistic (2).

---

[4] In principle, $M$ is also a poor explanation if $Z(M) \leq -5$. In this case the model explains the data too well! However, this is not a problem I face in this paper.

## Model #1

In the first and simplest model, there are four detection probabilities:

$pa_i$     Probability Alice detects a photon that arrives while she is performing measurement $i = 0,1$

$pb_k$     Probability Bob detects a photon that arrives while he is performing measurement $k = 0,1$

The model predicts the expected counts as:

(3a) $\qquad \hat{a}_{ij}^m = 2N \times pa_i \times qa_{ij}^m$

(3b) $\qquad \hat{b}_{kl}^m = 2N \times pb_k \times qb_{kl}^m$

(3c) $\qquad \hat{c}_{ijkl}^m = N \times pa_i \times pb_k \times qc_{ijkl}^m$

Here, $N$ is the expected number of photon pairs generated per *quadrant*, which is my term for a pair of settings $(i, k)$, one each for Alice and Bob. I assume that every quadrant in every experiment has the same expected number of pairs, which was surely Weihs' intention. The factor 2 in (3a) occurs because Alice's singles for setting $i$ include pairs generated in two quadrants, $(i, 0)$ and $(i, 1)$. The factor 2 in (3b) is explained similarly.

Because the detection probabilities are the same for both results for a given setting, the four predicted counts of coincidences in a quadrant are proportional to the theoretical probabilities, i.e., for each pair of settings $(i, k)$ and experiment $m$:

(4) $\qquad \dfrac{\hat{c}_{ijkl}^m}{\left(\hat{c}_{i0k0}^m + \hat{c}_{i1k0}^m + \hat{c}_{i0k1}^m + \hat{c}_{i1k1}^m\right)} = qc_{ijkl}^m$

This is the "fair sampling" assumption that most investigators make. It is implied by (3c), but not vice versa. For example, (4) could still hold if the number of photon pairs differed by quadrant.

The parameter values for which Model #1 fits the data best are as follows. The density matrix is:

**Table 1: Density Matrix for Model #1**

| 0.0153 | -0.0418+0.0003$i$ | 0.0317-0.0$i$ | -0.0026-0.0$i$ |
|---|---|---|---|
| -0.0418-0.0003$i$ | 0.4798 | -0.4341+0.0$i$ | -0.0388-0.0$i$ |
| 0.0317+0.0$i$ | -0.4341-0.0$i$ | 0.4867 | 0.0395-0.0003$i$ |
| -0.0026+0.0$i$ | -0.0388+0.0$i$ | 0.0395+0.0003$i$ | 0.0176 |

This is quite similar to the density matrix for the singlet state:

Table 2: Density Matrix for the Singlet State

| 0 | 0 | 0 | 0 |
|---|---|---|---|
| 0 | 0.5 | -0.5 | 0 |
| 0 | -0.5 | 0.5 | 0 |
| 0 | 0 | 0 | 0 |

The values of the remaining adjustable parameters are:

Table 3: Pairs per Quadrant and Detection Probabilities for Model #1

| N | $pa_0$ | $pa_1$ | $pb_0$ | $pb_1$ |
|---|---|---|---|---|
| 963,382 | 0.05110 | 0.05393 | 0.03657 | 0.03566 |

How well does this model fit the data?  The value of the X statistic is 22,054.07.  The model has 15 adjustable parameters in the density matrix (16 real numbers to define the real and imaginary parts of a self-adjoint matrix, less one to account for the constraint that the trace equals 1).[5]  The number of photon pairs per quadrant and four detection probabilities are also adjustable, for a total of 20 parameters.  Thus $DF = 984 - 20 = 964$.

Thus X is 480.31 standard deviations above its mean!

Figure 2 consists of 24 panels.  Each panel shows the actual counts (the black diamonds) from all 41 *scanblue* experiments for one of the 24 categories of counts, and compares them with predictions plus-or-minus a standard error (the three red lines) of Model #1.  The four panels at the top, comparing the predictions of Alice's unpaired singles to the observed counts, show the most clearly biased predictions.  If Alice's detection probabilities could differ by result as well as by setting, these biases could be greatly reduced.  And if I allow this license to Alice, can I do less for Bob?

## Model #2

Making this assumption I have:

$pa_{ij}$     Probability Alice detects a photon that arrives in result channel $j = 0,1$ while she is performing measurement $i = 0,1$

$pb_{kl}$     Probability Bob detects a photon that arrives in result channel $l = 0,1$ while he is performing measurement $k = 0,1$

This model predicts the expected counts to be:

(5a) $\quad\quad \hat{a}_{ij}^m = 2N \times pa_{ij} \times qa_{ij}^m$

---

[5] In principle some allowance should be made for the fact that $\rho$ is constrained to be positive semi-definite.  This would increase DF, and thus reduce the standard deviations by which X exceeds its mean.  But not by much.

(5b) $$\hat{b}_{kl}^m = 2N \times pb_{kl} \times qb_{kl}^m$$

(5c) $$\hat{c}_{ijkl}^m = N \times pa_{ij} \times pb_{kl} \times qc_{ijkl}^m$$

In this model, the four predicted counts of coincidences in a quadrant are no longer proportional to the theoretical probabilities.

The parameter values for which this model fits the data best are as follows. The density matrix is still rather similar to the singlet state:

**Table 4: Density Matrix for Model #2**

| 0.0180 | -0.0371+0.0$i$ | 0.0312-0.0$i$ | -0.0028-0.0002$i$ |
|---|---|---|---|
| -0.0371-0.0$i$ | 0.4782 | -0.4358+0.0$i$ | -0.0384-0.0$i$ |
| 0.0312+0.0$i$ | -0.4358-0.0$i$ | 0.4879 | 0.0469+0.0$i$ |
| -0.0028+0.0002$i$ | -0.0384+0.0$i$ | 0.0469-0.0$i$ | 0.0159 |

The number of photon pairs per quadrant is $N = 964{,}212$. The detection probabilities are:

**Table 5: Detection Probabilities for Model #2**

| $(i,j)$ or $(k,l)$ | (0,0) | (0,1) | (1,0) | (1,1) |
|---|---|---|---|---|
| $pa_{ij}$ | 0.04855 | 0.05344 | 0.05126 | 0.05638 |
| $pb_{kl}$ | 0.03627 | 0.03681 | 0.03655 | 0.03473 |

How well does Model #2 fit the data? The value of the *X* statistic is 3035.37. The model has 15 adjustable parameters in the density matrix, the number of photon pairs per quadrant, and eight detection probabilities, for a total of 24. Thus $DF = 984 - 24 = 960$.

Thus X is 47.36 standard deviations above its mean.

Both Adenier and Krennikov (2007) and De Raedt et al (2012) examine models very similar to this one. Adenier and Khrennikov concluded that "…this explicit use of fair sampling [their characterization of equations (5a-c)] cannot be maintained to be a reasonable assumption as it leads to an apparent violation of the no-signaling principle [a prediction of quantum theory]." De Raedt et al found that "[a]pparently, including [this] model for the detector efficiency does not resolve the conflict between the experimental data of Weihs et al and quantum theory of the EPRB thought experiment."

I agree that this model does not fit the *scanblue* data. But I don't agree that one must necessarily abandon the quantum theory of the EPRB thought experiment in order to make further progress. Instead, I further modify the filter by introducing yet more adjustable parameters. Figure 3 (same as Fig. 2 save that the predictions are those of Model #2) suggests what these new parameters should be. While Model #2's predictions of both Alice's and Bob's unpaired singles appear to be unbiased, there are still biases in some of the panels showing the coincidence categories.

## Model #3

The third model abandons the assumption that the probability of identifying a coincidence should equal the product of the probabilities of detecting the two photons. This assumption is made routinely (e.g., see Brunner et al, 2007; de Barros & Suppes, 2000; Massar et al, 2002; Massar & Pironio, 2003; Wilms et al, 2008; and, of course, Adenier & Khrennikov, 2007; De Raedt et al, 2012), because it seems natural that (a) detecting the two photons of a pair ought to be independent events, and (b) if one detects both photons of a pair, one has naturally identified that coincidence. Clearly, if either (a) and (b) or both are false, then the assumption that began this paragraph must fail. I explain later how (a) can be false. Here I explain how (b) can be false.

Coincidences are identified by forming pairs of detections, one each from Alice's and Bob's detection logs, and then accepting or rejecting each pair based on a test involving the two detection times. The test is, accept if $|t_b - (t_a - \delta)| \leq w$, reject otherwise, where $t_a, t_b$ are Alice and Bob's detection times, $\delta$ is a time offset meant to synchronize Alice's log with Bob's, and $w$ is the width (in ns) of the detection window. As shown in Bigelow (2009), depending on one's choice of $\delta$ and $w$, the test will accept some detection pairs that do not correspond to photon pairs (false positives), and will reject some detection pairs that do (false negatives). The fact that there are false negatives invalidates the assumption.

Moreover, the false negative rate depends on Alice's and Bob's settings and results. This is, I think, because the distribution of the delay between a photon's arrival and its detection depends on the observer's setting and result. See Bigelow (2009) and the discussion of Fig. 6 later in this paper. Also see Willeboordse (2008).

Model #3 has sixteen more adjustable parameters than Model #2.

$pa_{ij}$    Probability Alice detects a photon that arrives in result channel $j = 0,1$ while she is performing measurement $i = 0,1$

$pb_{kl}$    Probability Bob detects a photon that arrives in result channel $l = 0,1$ while he is performing measurement $k = 0,1$

$pc_{ijkl}$    Probability a coincidence is identified if a pair of photons arrives, one in Alice's result channel $j$ while she is performing a measurement $i$, and the other in Bob's result channel $l$ while he is performing a measurement $k$

Model #3 predicts the expected counts as:

(6a) $\quad \hat{a}_{ij}^m = 2N \times pa_{ij} \times qa_{ij}^m$

(6b) $\quad \hat{b}_{kl}^m = 2N \times pb_{kl} \times qb_{kl}^m$

(6c) $\quad \hat{c}_{ijkl}^m = N \times pc_{ijkl} \times qc_{ijkl}^m + \left(\hat{a}_{ij}^m \times \hat{b}_{kl}^m\right) \times (w^m/\text{T})$

The extra term $(\hat{a}_{ij}^m \times \hat{b}_{kl}^m) \times (w^m/\text{T})$ in (6c) is an estimate of false positives (i.e., pairs of detections that are identified as coincidences but do not correspond to entangled photon pairs—see Bigelow 2009). The parameter $w^m$ is the width of the detection window for experiment $m$,[6] while T = 5 sec or $5 \times 10^9$ ns is the common duration of all the *scanblue* experiments.

The parameter values for which Model #3 fits the data best are as follows. The density matrix is again much like the singlet state:

**Table 6: Density Matrix for Model #3**

| 0.0117 | -0.0384-0.0074$i$ | 0.0324-0.0055$i$ | 0.0032-0.0010$i$ |
|---|---|---|---|
| -0.0384+0.0074$i$ | 0.4851 | -0.4525+0.0823$i$ | -0.0399+0.0176$i$ |
| 0.0324+0.0055$i$ | -0.4525-0.0823$i$ | 0.4926 | 0.0486-0.0121$i$ |
| 0.0032+0.0010$i$ | -0.0399-0.0176$i$ | 0.0486+0.0121$i$ | 0.0106 |

In this model it is not possible to estimate the number of photon pairs per quadrant independently of the detection probabilities. Only the products can be estimated. They are:

**Table 7: Adjustable Parameters for Singles Detections, Model #3**

| $(i,j)$ or $(k,l)$ | (0,0) | (0,1) | (1,0) | (1,1) |
|---|---|---|---|---|
| $N \times pa_{ij}$ | 46,812.68 | 51,521.92 | 49,416.17 | 54,362.87 |
| $N \times pb_{kl}$ | 35,078.74 | 35,369.69 | 35,272.19 | 33,454.38 |

**Table 8: Adjustable Parameters for Coincidences, $N \times pc_{ijkl}$, Model #3**

| | | Bob's Setting & Result | | | |
|---|---|---|---|---|---|
| | | (0,0) | (0,1) | (1,0) | (1,1) |
| Alice's Setting & Result | (0,0) | 1448.14 | 1701.85 | 1540.10 | 1759.54 |
| | (0,1) | 1730.72 | 2005.93 | 1867.75 | 2071.06 |
| | (1,0) | 1621.77 | 1840.52 | 1704.22 | 1960.79 |
| | (1,1) | 1622.16 | 1858.88 | 1721.61 | 1957.92 |

How good is the fit? The value of the X statistic is 1689.95. The model has 15 adjustable parameters in the density matrix, and 24 detection probabilities are also adjustable, for a total of 39. Thus $DF = 984 - 39 = 945$.

So X is 17.14 standard deviations above its mean. By my chosen criterion, Model #3 does not fit the *scanblue* data.

Figure 4 (same as Figs 2 and 3 save that the predictions are those of Model #3) shows that the biases have been virtually eliminated from every panel.

---

[6] In his original analysis, Weihs used detection windows 5-6 ns wide for these experiments. I use a window about 30 ns wide. See Bigelow (2009) for a discussion of how I select the window, and why I think use of Weihs' windows yields unnecessarily high numbers of false negatives.

## Model #4

If there is a pattern to the remaining discrepancies, it has eluded me. I therefore treat them as noise. I suppose the parameters $pa_{ij}, pb_{kl}, pc_{ijkl}$ in (6a-c) differ among the experiments in an apparently random fashion. In Model #4, then, I give these parameters probability distributions whose means over the experiments will be the same as the values from Tables 7 and 8. I denote the coefficients of variation of the distributions by $cva_{ij}, cvb_{kl}, cvc_{ijkl}$. Then it is easy to show that, on the assumption that the model's predictions are correct, the variances of the counts are (see Appendix B):[7]

(7a) $$V(ua_{ij}^m) = \widehat{ua}_{ij}^m + \left(\widehat{ua}_{ij}^m \times cva_{ij}\right)^2$$

(7b) $$V(ub_{kl}^m) = \widehat{ub}_{kl}^m + \left(\widehat{ub}_{kl}^m \times cvb_{kl}\right)^2$$

(7c) $$V(c_{ijkl}^m) \approx \hat{c}_{ijkl}^m + \left(\hat{c}_{ijkl}^m \times cvc_{ijkl}\right)^2$$

So the variances in the X-statistic will be increased, and (2) must be replaced by:

(8) $$Xrev(M) = \sum_m \left[ \sum_{ij} \frac{\left(ua_{ij}^m - \widehat{ua}_{ij}^m\right)^2}{V(ua_{ij}^m)} + \sum_{kl} \frac{\left(ub_{kl}^m - \widehat{ub}_{kl}^m\right)^2}{V(ub_{kl}^m)} + \sum_{ijkl} \frac{\left(c_{ijkl}^m - \hat{c}_{ijkl}^m\right)^2}{V(c_{ijkl}^m)} \right]$$

Using the parameter values from Model #3, $X = 1689.95$. In order for Model #4 to fit the data we must have $Xrev \approx 945$. This can be done by increasing all the variances by a factor of about 1.8. Taking into account the relative sizes $\widehat{ua}_{ij}^m \sim 50{,}000$, $\widehat{ub}_{kl}^m \sim 35{,}000$, $\hat{c}_{ijkl}^m \sim 400$, the coefficients of variation that accomplish this are $cva_{ij} \sim 0.004$, $cvb_{kl} \sim 0.005$, $cvc_{ijkl} \sim 0.05$.

There is reason to think that $pa_{ij}, pb_{kl}, pc_{ijkl}$ could vary this much from one *scanblue* experiment to the next. The rather extended argument proceeds in the following steps.

***A period of 100 ns was built into Alice's and Bob's detection logs.***

Weihs (2007) reports that each observer chose his setting anew every 100 ns.

***The probability Alice (Bob) detects a photon depends on when in her (his) 100 ns cycle it arrives.***

The time to switch from one setting to the other was 14 ns, and detector pulses that occurred during the switching period were suppressed. Since the setting will be switched during half the 100 ns cycles, the detection probability during those 14 ns should be half as large as during the remaining 86 ns of the cycle.

I tested whether this phenomenon occurred in each of the *scanblue* experiments. I partitioned each 100 ns interval in Alice's (Bob's) detection logs into 100 bins indexed by $\alpha = 1, \ldots, 100$ ($\beta = 1, \ldots, 100$). I must make do by assigning detections to bins rather than photon arrivals. Figure 5 shows the results for experiment *scanblue110*.

---

[7] Equation (7c) is approximate because it ignores the effect of false positive counts on the variance.

The switching interval stands out, though it seems to be longer than 14 ns. But much more is going on. Most startling is a very prominent 20 ns cycle for detections at Alice's setting $i = 1$ and Bob's setting $k = 1$. These cycles need not be due entirely to variations in detection probabilities over bins. As I show later, they are likely due in part to variations in the delay between a photon's arrival and its detection.

I have carried out the same analysis for all 41 experiments in the *scanblue* series. It is quite startling that if the bins are shifted by an appropriate multiple of 20 ns, all 41 figures are virtually identical! (Once shifted, the detection counts as a function of bin from any two experiments have a correlation greater than 0.95.) A different shift may be needed for Alice than for Bob. I will refer to this adjustment later as *reconciling the zero times*.[8]

### *Argument that $pa_{ij}$ and $pb_{kl}$ can vary across experiments*

Fig. 5 is strong evidence that detection probabilities vary within each experiment, and the fact that they vary at all suggests to me that they could vary from one experiment to another. True, the correlations between experiments are high, which should make the between-experiment differences small. But I am only looking for small differences—coefficients of variation of 0.004 (0.005) for Alice's (Bob's) detection probabilities.

### *If you know the arrival time modulo 100 ns of Alice's photon, you know Bob's*

Let $\lambda(\alpha, \beta)$ be the probability that when the central device generates a pair of photons, Alice receives hers in bin $\alpha$ and Bob receives his in bin $\beta$. The arrivals of the two photons should be separated by an approximately constant interval, so that $\lambda(\alpha, \beta)$ should be zero unless $(\beta - \alpha)\ modulo\ 100$ is close to some offset $\delta$.

To check this, I assign coincidences to pairs of bins, one for Alice's detection, one for Bob's. As shown in Fig. 6, the coincidence counts are highly concentrated along a diagonal—not the main diagonal, but a diagonal above or below the main diagonal by an amount equal to the offset $\delta$ required to synchronize Alice's detection log with Bob's. In this experiment (*scanblue110*) the offset is about 15 ns.

In the plots of coincidences with Alice's setting $i = 1$ and/or Bob's setting $k = 1$, some of the points appear to have been swept off the diagonal. It seems logical that the coincidences below the diagonal were moved there by a delay in Alice's detection, while the coincidences above the diagonal were moved there by a delay in Bob's detection. In Bigelow (2009) I suggested that such delays occurred.

---

[8] Alice's and Bob's detection logs both start over at zero in each of the 41 *scanblue* experiments. But their clocks simply keep running and the 100 ns cycle keeps repeating. Shifting Alice's and Bob's recorded detection times as described in the text causes the new zero time of each experiment fall at the same point in the 100 ns cycle. Hence the name, reconciling the zero times.

### *Alice's and Bob's detections need not be independent*

Denote the detection probabilities by bin as $pa_{ij}(\alpha)$ for Alice and $pb_{kl}(\beta)$ for Bob. I assume photons arriving at Alice's (Bob's) station are distributed uniformly over bins, so the average detection probabilities for Alice and Bob will be:

$$\overline{pa}_{ij} = \frac{1}{100} \sum_\alpha pa_{ij}(\alpha) \qquad \overline{pb}_{kl} = \frac{1}{100} \sum_\beta pb_{kl}(\beta)$$

The probability that both photons in a pair are detected will be:

(9) $$\overline{pab}_{ijkl} = \sum_{\alpha,\beta} pa_{ij}(\alpha) \times pb_{kl}(\beta) \times \lambda(\alpha,\beta)$$

Since $\lambda(\alpha,\beta)$ is not uniform, $\overline{pab}_{ijkl}$ can be less than or greater than, and certainly needn't be equal to, $\overline{pa}_{ij} \times \overline{pb}_{kl}$.

### *Different experiments have different distributions $\lambda(\alpha,\beta)$*

Figure 7 shows that the distribution $\lambda(\alpha,\beta)$ shifts by about 25 bins in the 460 seconds elapsed between experiments *scanblue110* and *scanblue151*. In experiment *scanblue110*, coincidences cluster around a time difference $(t_b - t_a)$ of about 15 ns. In experiment *scanblue151* coincidences cluster at a time difference of about -10 ns, once I have reconciled the zero times of the two experiments. I attribute this to Alice's clock running faster than Bob's by about 0.055 ns per second. [9]

How much might $\overline{pab}_{ijkl}$ differ from $\overline{pa}_{ij} \times \overline{pb}_{kl}$? Suppose that $\lambda(\alpha,\beta)$ equals 0.01 if $\beta - \alpha = \delta$ for a specified offset $\delta$, and is zero otherwise. In Figure 8 I have calculated the ratio $\overline{pab}_{ijkl}/(\overline{pa}_{ij} \times \overline{pb}_{kl})$ as a function of this offset, taking the singles distributions over bins from Fig. 5 as my functions $pa_{ij}(\alpha)$ and $pb_{kl}(\beta)$. [10] The red regions on the horizontal axis show the range of offsets for the 41 *scanblue* experiments. The smallest coefficient of variation among the ratios $\overline{pab}_{ijkl}/(\overline{pa}_{ij} \times \overline{pb}_{kl})$ is 0.021, for $(i,j,k,l) = (0,1,0,0)$. The largest is 0.089, for $(i,j,k,l) = (1,1,1,1)$.

### *Argument that $pc_{ijkl}$ can vary across experiments*

I express $pc_{ijkl}$ as:

(10) $$pc_{ijkl} = pa_{ij} \times pb_{kl} \times \left[\frac{pab_{ijkl}}{pa_{ij} \times pb_{kl}}\right] \times \left[\frac{pc_{ijkl}}{pab_{ijkl}}\right]$$

---

[9] This dependence of $\lambda(\alpha,\beta)$ on the drift of Alice's clock relative to Bob's would not occur if Alice and Bob had switching cycles of incommensurate lengths, say Alice 100 ns and Bob 137 ns. This is something to consider if these experiments are ever repeated.

[10] As mentioned earlier, I believe the variability in Fig. 5 is due in part to delays between the arrival and detection of photons, and in part to variation of the detection probability over bin. This calculation assumes it is all due to variations in detection probability. But I intend this calculation only for illustration.

Earlier I argued that $pa_{ij}$ and $pb_{kl}$ could vary from one experiment to the next. The previous section shows that the ratio $pab_{ijkl}/(pa_{ij} \times pb_{kl})$ can vary as well, where $pab_{ijkl}$ is the joint probability of detecting both photons of a pair. When transitioning from Model #2 to Model #3, I argued that $pc_{ijkl}$ could differ from $pab_{ijkl}$. The coefficient of variation of the ratio $pab_{ijkl}/(pa_{ij} \times pb_{kl})$ alone is nearly as large as I am looking for ($cvc_{ijkl} \sim 0.05$). When the variations in $pa_{ij}, pb_{kl}$, and the ratio $pc_{ijkl}/pab_{ijkl}$ are considered, surely the overall coefficient of variation cannot fall far short of the needed value.

## CONCLUSIONS

Based on the foregoing analysis, I have formed two conclusions.

- Contrary to the conclusions of Adenier and Khrennikov and of De Raedt et al, the *scanblue* data are consistent with quantum theory.

- The EPRB experiments of Weihs et al do not approximate the ideal experiment—the one they really wanted to do—as closely as most researchers in this area think they do. The real experiments suffer from more noise and distortion that is usually assumed.

The quotation from Bell's "Bertlemann's Socks" paper can serve as a preamble to my argument:

> "It is true that practical experiments fall far short of the ideal, because of counter inefficiencies, or analyzer inefficiencies, or geometrical imperfections, and so on. It is only with added assumptions, or conventional allowance for inefficiencies and extrapolation from the real to the ideal, that one can say the inequality is violated. Although there is an escape route here, it is hard for me to believe that quantum mechanics works so nicely for inefficient practical set-ups and is yet going to fail badly when sufficient refinements are made." (Bell, 1981)

Contemplate a hypothetical sequence of EPRB experiments, from the "inefficient practical set-ups" of Bell's day to vastly more refined experiments of the future. The ideal that Bell refers to is the limit of this sequence. This limit might be the EPRB thought experiment, or it might be something else, whether locally causal and realistic or weirder than quantum theory.

The conclusions one can draw from actual EPRB experiments depends on how closely the real experiments approach the limiting ideal. Bell said the experiments of his time fell "far short of the ideal," but he also said that "quantum theory works so nicely for inefficient practical set-ups." So which is it? Are practical EPRB experiments such as those of Weihs et al far from the ideal or close to it?

From my perusal of the literature I judge that most physicists working in this area would answer this question "both!" There is general agreement that practical experiments have gaping loopholes (low detection rates in photon experiments is the prime example). But there is also widespread (though not universal) agreement that current experiments satisfy (1) a "fair sampling" assumption, and (2) an assumption that counts of detections and coincidences have Poisson distributions. Together these two assumptions imply that current real experiments are "close to" ideal experiments in the sense that the results obtained from the real experiment are related in a simple and straightforward way to the results one would obtain from the ideal experiment. In particular, the real results are greatly attenuated but not much distorted compared to the ideal results.

Both Adenier & Khrennikov and De Raedt et al accept these two assumptions. Adenier & Khrennikov analyzed the same data I did, and they concluded that the data were inconsistent with both quantum and classical mechanics. De Raedt et al analyzed data from 23 of the experiments of Weihs et al—not the *scanblue* experiments, but very similar. They concluded that the data were inconsistent with quantum theory, but proposed a locally-causal model consistent with classical physics that they believe can explain the Weihs data.[11]

Both of these papers show, in effect, that one cannot simultaneously hold that these assumptions are true *and* that the data of Weihs et al derive from an EPRB thought experiment. In both papers the choice is made to give up the EPRB explanation for the data, rather than abandon fair sampling and Poisson errors. But I have described some quite ordinary mechanisms that could account for some degree of failure of the fair sampling assumption, and I have discussed why the customary statistical analysis is likely to underestimate the standard errors of the estimated counts. So it seemed to me every bit as reasonable to retain EPRB and to relax the assumptions as I moved from Model #1 to #2 to #3 to #4. This amounted to supposing that the experiments of Weihs et al are farther from Bell's ideal experiment than is generally believed. And once I relaxed the assumptions, I was able tell a plausible story (my Model #4) for how the *scanblue* data could derive from an EPRB thought experiment.

This does not rule out other derivations of the *scanblue* data. It would be interesting to see if a model built around a paradigm other than an EPRB thought experiment can fit the data as well as Model #4, or perhaps even better. De Raedt et al propose a specific local-realist model to replace the EPRB thought experiment, and there is no shortage of other local-realist candidates in the literature (e.g., Gisin & Gisin, 1999; Hofer, 2001a & 2001b; Larssen, 1999; Szabo & Fine, 2002; Thompson & Holstein, 2002). But this will have to be deferred to another paper.

---

[11] Most earlier authors seem to have concluded the reverse, that (almost) all EPRB experiments confirmed quantum mechanics (e.g., see Aspect, 2002). But there are statistics one can calculate from the data of Weihs et al that clearly conflict with predictions of quantum theory, if one accepts the two assumptions given in the text. I have had no opportunity to see whether the same is true of the many other EPRB experiments listed in Aspect (2002).

de Barros JA, Suppes P (2000), *Inequalities for dealing with detector inefficiencies in Greenberger-Horne-Zeilinger-type experiments*, Phys Rev Lett 84, 793-797. As of February 17, 2008: arXiv:quant-ph/0001034v1.

De Raedt H, De Raedt K, Michielsen K, Keimpema K, Miyashita S (2007), *Event-by-event simulation of quantum phenomena: Application to Einstein-Podolshy-Rosen-Bohm experiments*, J Comp Theor Nanosci 4: 957-991. As of February 25, 2008, arXiv:quant-ph/0712.3781v2.

De Raedt H, Michielsen K, Jin F (2012), *Einstein-Podolsky-Rosen-Bohm laboratory experiments: Data analysis and simulation*, Proceedings of the International Conference on Advances in Quantum Theory, FPP6 – Foundations of Probability and Physics 6, A. Khrennikov et al., eds. As of December 13, 2011: arXiv:quant-ph/1112.2629v1.

Einstein A, Podolsky B, Rosen N (1935), *Can quantum-mechanical description of physical reality be considered complete?*, Physical Review 47(10):777-780.

Gisin N, Gisin B (1999), *A local hidden variable model of quantum correlation exploiting the detection loophole*, Phys Lett A 260: 323-327. As of January 14, 2008: arXiv:quant-ph/9905018v1.

Hofer WA (2001a), *Simulation of Einstein-Podolsky-Rosen experiments in a local hidden variables model with limited efficiency and coherence*. As of January 21, 2008: arXiv:quant-ph/0103014v2.

Hofer WA (2001b), *Numerical simulation of Einstein-Podolsky-Rosen experiments in a local hidden variables model*. As of January 21, 2008: arXiv:quant-ph/0108141v1.

Larsson J-A (1999), *Modeling the singlet state with local variables*, Phys Lett A 256: 245-252. As of October 31, 2011: arXiv:quant-ph/9901074v1.

Massar S, Pironio S, Roland J, Gisin B (2002), *A zoology of Bell inequalities resistant to detector inefficiency*, Phys Rev A 66, 052112. As of January 15, 2008: arXiv:quant-ph/0205130v2.

Massar S, Pironio S (2003), *Violation of local realism vs detector effiiciency*, Phys Rev A 68, 062109. As of January 15, 2008: arXiv:quant-ph/0210103v2.

Peres A (1999), *All the Bell inequalities*, Foundations of Physics 29, 589-614. As of January 16, 2008: arXiv:quant-ph/9807017v2.

Szabo LE, Fine A (2002), *A local hidden variable theory for the GHZ experiment*, Phys Lett A 295, 229-240. As of December 20, 2007: arXiv:quant-ph/0007102v7.

Thompson CH, Holstein H (2002), *The "chaotic ball" model: local realist and the Bell test "detection loophole"*. As of December 13, 2007, arXiv:quant-ph/0210150v3.
17

Weihs G (2007), *A test of Bell's inequality with spacelike separation*, in Foundation of Physics and Probability – 4, AIP Conference Proceedings 889, pp. 250-260.

Weihs G, Jennewein T, Simon C, Weinfurter H, Zeilinger A (1998), *Violation of Bell's inequality under strict Einstein locality conditions*, Phys Rev Lett 81, 5039-5043. As of December 22, 2007, arXiv:quant-ph/9810080v1.

Willeboordse FH (2008), *Bell correlations and equal time measurements*. As of April 9, 2008, arXiv:quant-ph/0804.0721v2.

Wilms J, Disser Y, Alber G, Percival IC (2008), *Locality, detection efficiencies, and probability polytopes*. As of April 23, 2009, arXiv:quant-ph/0808.2126v1.

## APPENDIX A: OPERATORS IN THE *SCANBLUE* EXPERIMENTS

Measurements by Alice or Bob are modeled as 2×2 self-adjoint linear operators, which can all be expressed as linear combinations of the identity plus the three Pauli matrices:

$$I = \begin{bmatrix} 1 & \\ & 1 \end{bmatrix} \qquad \sigma_x = \begin{bmatrix} & 1 \\ 1 & \end{bmatrix} \qquad \sigma_y = \begin{bmatrix} & -i \\ i & \end{bmatrix} \qquad \sigma_z = \begin{bmatrix} 1 & \\ & -1 \end{bmatrix}$$

In each experiment Alice and Bob each switch randomly between two measurement settings that can be expressed as unit vectors. Bob's two vectors are $\vec{b}^0 = (0,0,1)$ and $\vec{b}^1 = (-1,0,0)$. These are largely arbitrary. The only necessity is that the two directions are 90° apart. Define the vector $\vec{\sigma} = (\sigma_x, \sigma_y, \sigma_z)$. Then the formulas for Bob's operators are:

$$B_{k0}^m - B_{k1}^m = \langle \vec{b}^k \cdot \vec{\sigma} \rangle$$
$$B_{k0}^m + B_{k1}^m = I$$

This leads to:

$$B_{00}^m = \tfrac{1}{2}(I + \sigma_z). \qquad B_{01}^m = \tfrac{1}{2}(I - \sigma_z).$$
$$B_{10}^m = \tfrac{1}{2}(I - \sigma_x). \qquad B_{11}^m = \tfrac{1}{2}(I + \sigma_x).$$

Alice's vectors are $\vec{a}^0 = (sin\theta, 0, cos\theta)$ and $\vec{a}^1 = (sin(\theta - \pi/2), 0, cos(\theta - \pi/2))$, where the angle $\theta$ differs from one experiment to the next as shown in Table A.1. The angle is proportional to the input voltage Alice applies to her box. That voltage consists of one or the other of the same two voltages that Bob uses, plus a bias that is constant throughout an experiment but varies from one experiment to the next. The bias runs from -100 in experiment scanblue110 to +95 in scanblue151. The documentation (Weihs, 2007) does not specify the units.

I build Alice's operators using the same equations as for Bob.

$$A_{i0}^m - A_{i1}^m = \langle \vec{a}^i \cdot \vec{\sigma} \rangle$$
$$A_{i0}^m + A_{i1}^m = I$$

**Table A.1: Alice's Bias Angles in the *scanblue* Series of Experiments**

| Experiment | $\theta$ | Experiment | $\theta$ | Experiment | $\theta$ |
|---|---|---|---|---|---|
| scanblue110 | $-1.00\pi$ | scanblue124 | $-0.30\pi$ | scanblue139 | $0.35\pi$ |
| scanblue111 | $-0.95\pi$ | scanblue125 | $-0.25\pi$ | scanblue140 | $0.40\pi$ |
| scanblue112 | $-0.90\pi$ | scanblue126 | $-0.20\pi$ | scanblue141 | $0.45\pi$ |
| scanblue113 | $-0.85\pi$ | scanblue127 | $-0.15\pi$ | scanblue142 | $0.50\pi$ |
| scanblue114 | $-0.80\pi$ | scanblue128 | $-0.10\pi$ | scanblue143 | $0.55\pi$ |
| scanblue115 | $-0.75\pi$ | scanblue129 | $-0.05\pi$ | scanblue144 | $0.60\pi$ |
| scanblue116 | $-0.70\pi$ | scanblue130 | $0$ | scanblue145 | $0.65\pi$ |
| scanblue117 | $-0.65\pi$ | scanblue131 | $0.05\pi$ | scanblue146 | $0.70\pi$ |
| scanblue118 | $-0.60\pi$ | scanblue132 | $0.10\pi$ | scanblue147 | $0.75\pi$ |
| scanblue119 | $-0.55\pi$ | scanblue133 | $0.15\pi$ | scanblue148 | $0.80\pi$ |
| scanblue120 | $-0.50\pi$ | scanblue134 | $0.20\pi$ | scanblue149 | $0.85\pi$ |
| scanblue121 | $-0.45\pi$ | scanblue135 | $0.25\pi$ | scanblue150 | $0.90\pi$ |
| scanblue122 | $-0.40\pi$ | scanblue136 | $0.30\pi$ | scanblue151 | $0.95\pi$ |
| scanblue123 | $-0.35\pi$ | scanblue137 | $0.35\pi$ | | |

## APPENDIX B: EQUATIONS FOR VARIANCES OF COUNTS

Here I derive equations (7a-c) for the variances of detection counts when both the number of events to be counted and the detection probability are random variables. Let:

$N$      Number of events available to be counted[12]

$x$      Probability of detecting an event

$n$      Number of events counted

On the assumption that detections are mutually independent, it is well known that:

(B.1)      $E(n|N,x) = Nx$

(B.2)      $V(n|N,x) = Nx(1-x)$

(Read $E(n|N,x)$ as "expected value of $n$ given $N$ and $x$," and similarly for $V(n|N,x)$.) I wish to integrate out the dependence on $N$ and $x$, but I can't do this to $V(n|N,x)$ directly because it is the second moment around the mean $E(n|N,x)$, and the mean depends on $N$ and $x$. Instead, I work with the second moment of $n$ around zero (any fixed point will do):

---

[12] Not to be confused with the use of $N$ as the expected number of photon pairs per quadrant in the main text.

(B.3) $$E(n^2|N,x) = V(n|N,x) + E(n|N,x)^2$$
$$= Nx - Nx^2 + N^2x^2$$

Now let $N$ and $x$ be independent random variables. I calculate the unconditional expectations of $n$ and $n^2$ to be:

(B.4) $$E(n) = E(N)E(x)$$

(B.5) $$E(n^2) = E(N)E(x) - E(N)E(x^2) + E(N^2)E(x^2)$$

If I express the second moments in (B.5) in terms of variances I have:

(B.6) $$V(n) = E(N)E(x) - E(N)V(x) - E(N)E(x)^2 + V(N)V(x)$$
$$+ V(N)E(x)^2 + E(N)^2 V(x)$$

Now specialize (B.6) for the case that $N$ has a Poisson distribution, so that $V(N) = E(N)$:

(B.7) $$V(n) = E(N)E(x) + E(N)^2 V(x)$$

Finally, substitute (B.4) into (B.7), and express the variance of $x$ in terms of its mean and coefficient of variation:

(B.8) $$V(n) = E(n) + \big(E(n)CV(x)\big)^2$$

# FIGURES

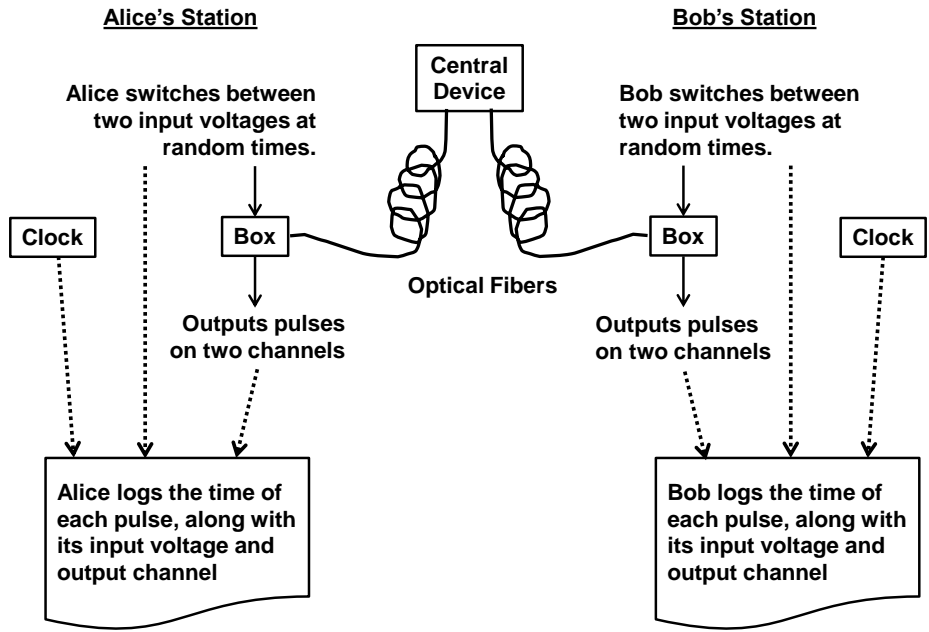

**Figure 1: The Experimental Set-up**

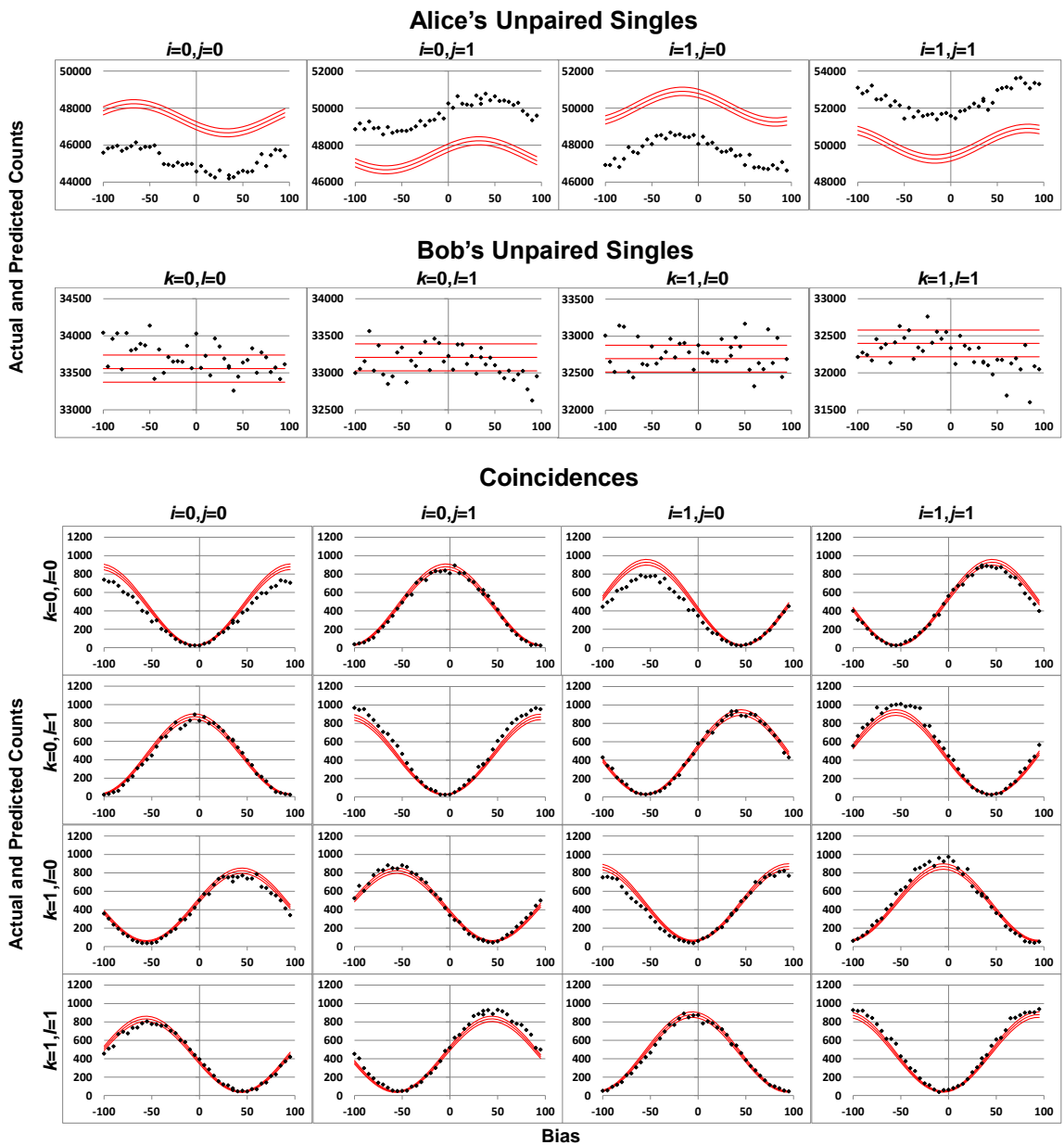

**Figure 2: Comparison of Observed Counts to Predictions of Model #1 ± One Standard Error**
(**Experiments** *scanblue110 - scanblue151*)

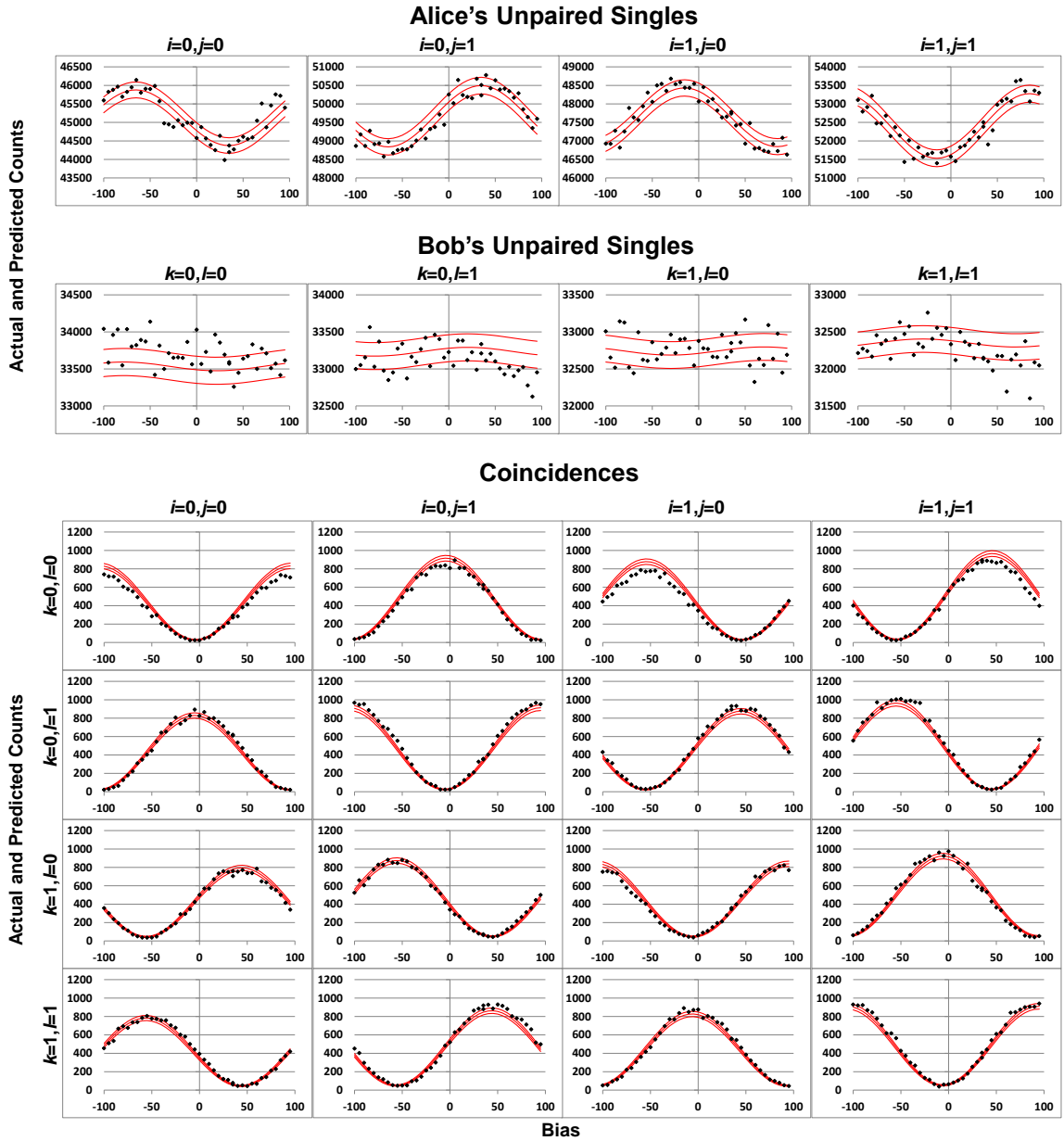

**Figure 3: Comparison of Observed Counts to Predictions of Model #2 ± One Standard Error**
(**Experiments** *scanblue110 - scanblue151*)

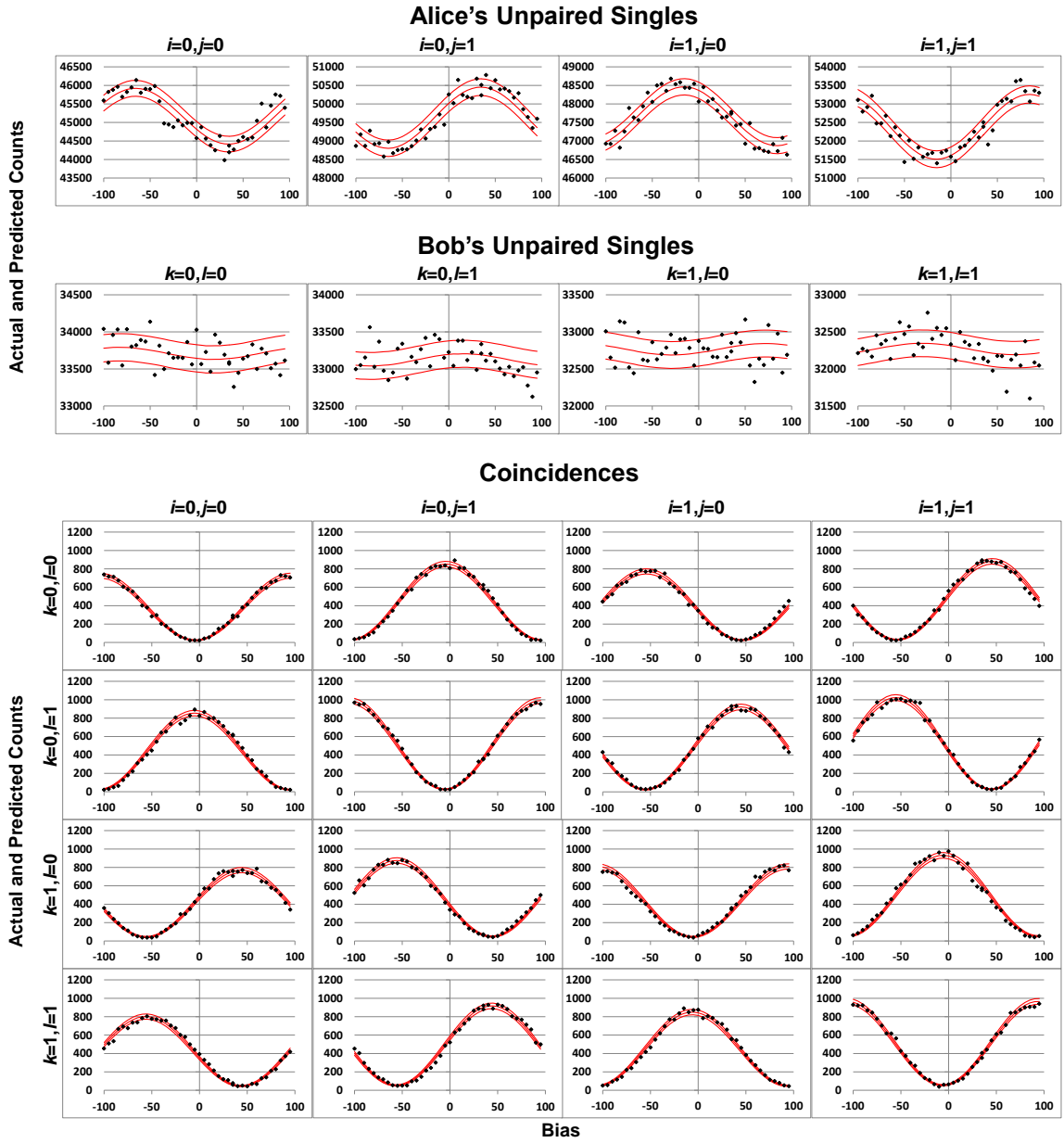

**Figure 4: Comparison of Observed Counts to Predictions of Model #3 ± One Standard Error**
(Experiments *scanblue110 - scanblue151*)

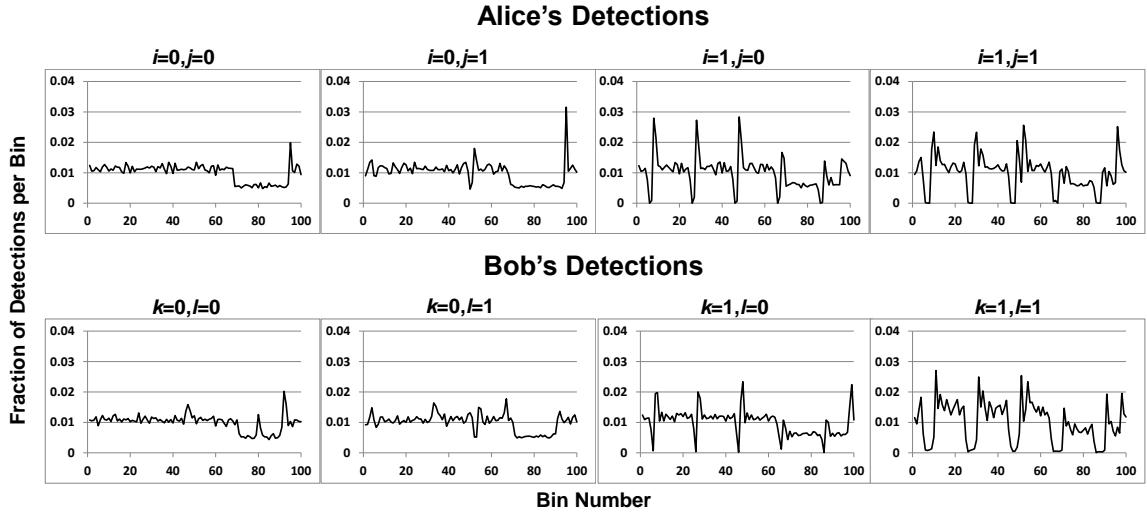

**Figure 5: Distribution of Detections Over 100 ns Segments**

**(Experiment scanblue110)**

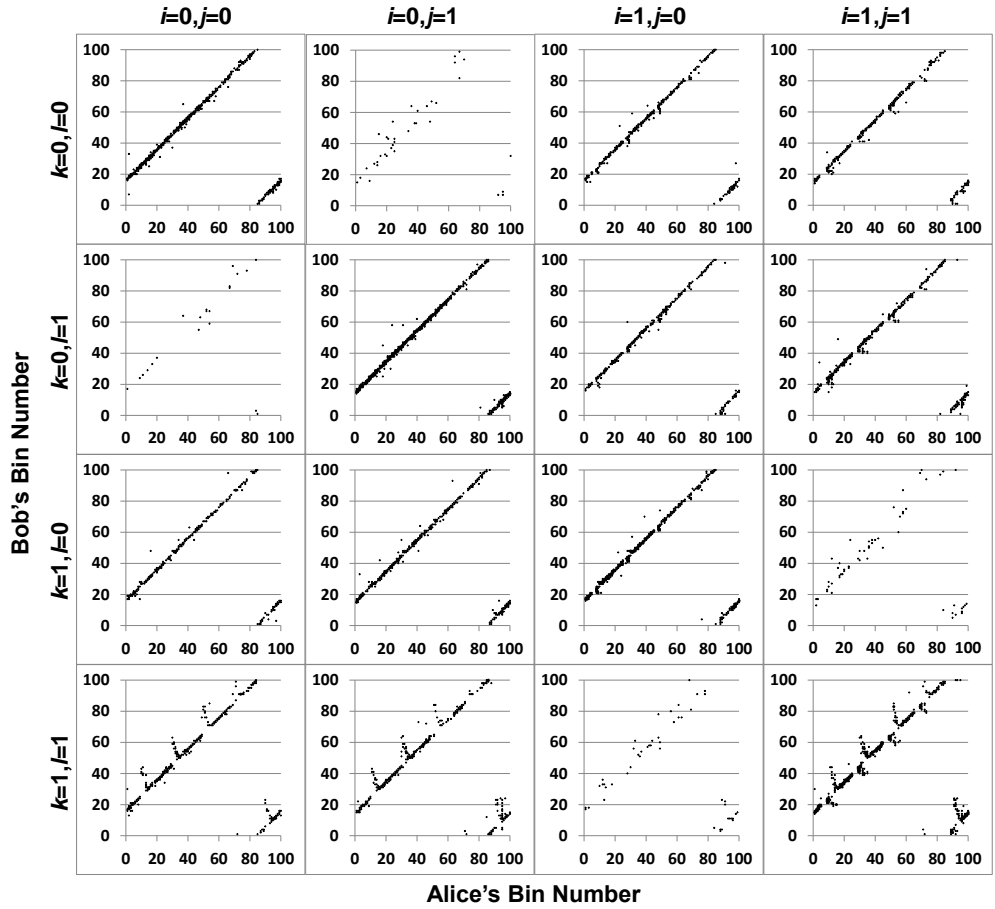

**Figure 6: Distribution of Coincidences**

**(Experiment scanblue110)**

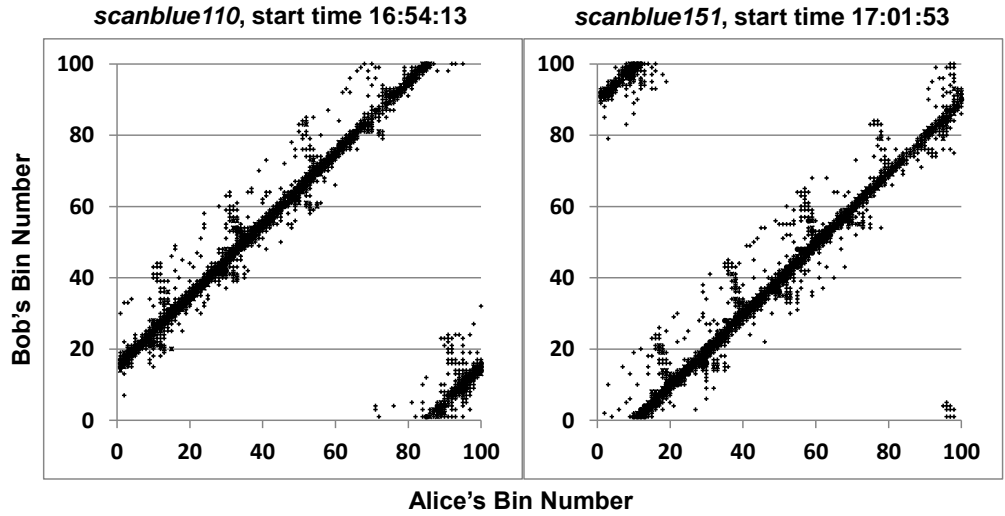

**Figure 7: Alice's Clock Runs Faster than Bob's**

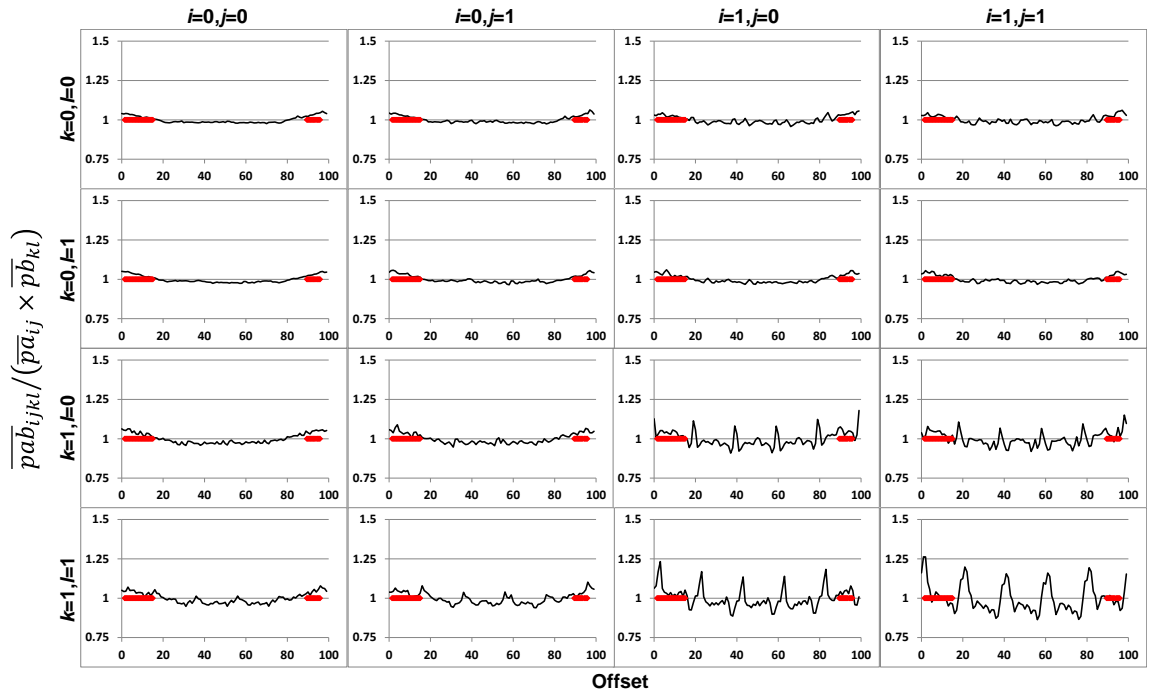

**Figure 8: The joint probability that both photons of a coincidence are detected deviates from the product of the marginal probabilities**